\documentclass[a4paper, 11pt, fleqn]{article}
\topmargin
-2cm

\usepackage{longtable}
\usepackage{graphicx}
\usepackage{cmap}					
\usepackage{mathtext} 				
\usepackage[T2A]{fontenc}			
\usepackage[utf8]{inputenc}			
\usepackage[english]{babel}	        

\begin{document}

\title{\bf{Surprising Variability of the Planetary Nebula IC~4997=QV~Sge}}

\author{V.P. Arkhipova, M.A. Burlak\footnote{E-mail:
marina.burlak@gmail.com}, N.P. Ikonnikova, G.V. Komissarova, \\V.F. Esipov, V.I. Shenavrin}

\date{\it{Sternberg Astronomical Institute,\\ Moscow State University (SAI MSU), Universitetskii pr. 13, Moscow, 119992 Russia}}

\renewcommand{\abstractname}{ }

\maketitle

\begin{abstract}
We present the results of a new epoch (2009-2019) of a long-term (50 years) photometric monitoring of the variable planetary nebula IC~4997 (QV~Sge). The integral (star + nebula) $UBV$ light curves display a continuing brightening of 0.$^{m}15$ in $V$, a slight rise (<0.$^{m}1$) in $B$, and constancy in $U$. The $B-V$ color has got redder from 0.$^{m}4$ in 2000 to 0.$^{m}7$ in 2019, whereas the $U-B$ color has not changed significantly at that time. We carried out near infrared (IR) $JHKL$ photometry in 2019, and comparing it to the data obtained in 1999-2006 we found the source to be fainter by 0.$^{m}4$ in $L$ and bluer by 0.$^{m}2$ in the $K-L$ color. The long-term brightness variations in the optical and IR regions are shown to be due mostly to the changing input of emission lines to the integral light.    

Low-dispersion spectroscopic observations carried out in 2010-2019 revealed a continuing decrease in the [OIII] $\lambda$4363 to H$\gamma$ intensity ratio: it decreased by a factor of $\sim3$ in 30 years and reached the level of 1960-1970. We discovered that the absolute intensities of [OIII] $\lambda$4959 и $\lambda$5007 nebular lines had increased by a factor of $>2$ from 1990 to 2019, whereas the [OIII] $\lambda$4363 auroral line had weakened by a factor of 2 comparing to the maximum value observed in 2000. The variation of H$\beta$ absolute intensity in 1960-2019 was shown to be similar to that of [OIII] $\lambda$4959 (and $\lambda$5007), but of smaller amplitude. The electron density in the outer part of the nebula was estimated from the [SII] and [ClIII] lines. Basing on the data on absolute intensities for the H$\beta$, [OIII] $\lambda4363, 4959$ lines and their ratios we propose a possible scenario describing the change of physical conditions ($N_e,T_e$) in IC~4997 in 1970-2019. The main features of spectral variability of IC~4997 could be explained by a variation of electron temperature in the nebula caused by not so much the change in ionizing flux from the central star as the variable stellar wind and related processes. The photometric and spectral changes observed for IC~4997 in 1960-2019 may be interpreted as an observable consequence of a single episode of enhanced mass loss from the variable central star.

\end{abstract}
 

\newpage

\section*{INTRODUCTION}

IC~4997 attracted particular attention as the first confirmed variable source of radiation among the young planetary nebulae (PNs). Liller and Aller (1957) reported a noticeable change in the $\lambda$4363 [OIII] to H$\gamma$ intensity ratio, having compared their own spectral measurements of 1956 with those of Menzel et al. (1941) made in 1938. Vorontsov-Velyaminov (1960) confirmed the variability of the ratio basing on the spectra obtained in Crimea in 1959-1960.   

IC~4997 received a great deal of interest in the 1960-1970s: the spectrophotometric observations were carried out by O'Dell (1963), Aller and Kaler (1964), Aller and Liller (1966), - and later by Ferland (1979), Feibelman et al. (1979), Purgathofer (1981). The optical spectrum of IC~4997 was the most thoroughly investigated in Hyung et al. (1994), Hyung and Aller (1993), where the relative intensities of more than 500 emission lines in the $\lambda\lambda3647-10049$ region were measured in the echelle spectra obtained in 1990 and 1991. The authors also noted the change in relative fluxes for quite a number of other emission lines besides [OIII] $\lambda$4363 and H$\gamma$ over a year of high-dispersion observations.
 
A large infrared (IR) excess in the 11~$\mu$m range was first detected for IC~4997 by Gillett et al. (1971). Natta and Panagia (1981), Pottasch et al. (1984), Lenzuni et al. (1989) thoroughly studied the IR spectrum of the nebula and its dust envelope.  Basing on IR photometry carried out in 1999-2006 Taranova and Shenavrin (2007) found the variability of IC~4997 with a peak-to-peak amplitude of 0.$^m$20--0.$^m$25 in $H$ and 0.$^m$05 in $J$. 

Radio observations carried out by Miranda et al. (1996), Miranda and Torrelles (1998) made it possible to construct for IC~4997 a map of 3.6~cm and 2~cm continuum radiation with angular resolution better than $0.^{\prime\prime}1$. The authors described new details in the structure of its outer and inner shells and confirmed the double-shell model for the nebula suggested previously by Hyung et al. (1994). Investigating the variability of IC~4997 radio flux Miranda and Torrelles (1998) found short timescale ($<1$~year) morphological changes in the inner shell structure in the vicinity of the central star $<0.3^{\prime\prime}$ and explained them by the collimated stellar wind impinging on the outer shell of the nebula. According to the archival data, the 5~GHz radio flux from the optically thin part of the nebula has decreased from 100~mJy around 1989 (Acker et al., 1992) to 45~mJy in 1996 (G\'{o}mez et al., 2002). Subsequent observations (Casassus et al., 2007, Pazderska et al., 2009) showed an increase in radio flux from 80 to 108~mJy at a frequency of 30~GHz but the data was not enough to analyze the radio spectrum in detail.  

The central star of IC~4997, HD 193538=QV Sge, was classified in a number of papers as a weak emission line star -- {\it wels}. It's hard to divide the observed integral optical spectrum into the components belonging to the nebula and the central star, but the CIV $\lambda$5801, 5812 and CIII $\lambda$4650 emission lines are identified in the echelle spectra with confidence and assigned to the star regarding their width (Hyung et al., 1994, Marcolino and de Ara\'{u}jo, 2003). Marcolino et al. (2007) noted that the CIV $\lambda$1549 emission line was not in a P~Cyg feature in the SWP 31683 spectrum (1987-09-01) taken from the IUE archive but analyzing an earlier IUE spectrum SWP 08578 (1980-03-27) they could see some difference: a slightly higher continuum and the NV $\lambda$1238 line possibly in a P~Cyg feature in the earlier spectrum. The authors concluded that such profiles are formed in the stellar wind of the central star and not in the nebula. This fact may imply the variability of the IC~4997 central star related to the existence of a variable stellar wind.  

According to the Sahai et al. (2011) morphological classification system, IC~4997 was assigned to bipolar PN regarding its Hubble Space Telescope image. The bright part of the nebula extends to $<2^{\prime\prime}$ from the central star and consists of two pairs of diametrically opposed lobes whose axes are perpendicular one to another.

Regular $UBV$ photometric and spectral observations of IC~4997 at the Crimean astronomical station (CAS) of SAI MSU were initiated by E.B.~Kostyakova and started in 1968. The angular visible size of the nebula is about $2^{\prime\prime}$. Both photometric and spectral observations realized as part of the program include the whole nebula together with the central star HD~193538. The results of photometric and spectral monitoring in different epochs were published in a quite number of papers: Vorontsov-Velyaminov et al. (1970), Kostyakova (1971), Kostyakova et al. (1973), Arkhipova et al. (1994), Kostyakova (1990, 1999), Kostyakova and Arkhipova (2009), Burlak and Esipov (2010). We continue observing IC~4997 on the CAS telescopes and in this paper we analyze the measurements obtained in 2009-2019 together with previous data.   

\section*{OBSERVATIONS}
\subsection*{$UBV$ photometry}

Although there is a good deal of single integral brightness estimates for IC~4997 published, it would hardly make sense to compare them as there are many strong emissions and the photometric systems are slightly different, which leads to a significant scatter in data. We succeeded to maintain a long-term monitoring at the same telescope with permanent equipment.

Our $UBV$ observations of IC~4997 have been carried out with the photon counting photometer constructed by Lyutyi (Lyutyi, 1971) mounted at the Cassegrain focus of the Zeiss-600 telescope of CAS since 1971. Previous results obtained mainly by Kostyakova~E.B. can be found in Kostyakova et al. (1973), Kostyakova (1991), Arkhipova et al. (1994), Kostyakova and Arkhipova (2009). In this paper we present recent $UBV$ photometry obtained in 2009-2019. As usual, HD~355464 was used as a reference star, and its magnitudes ($V=9.^{m}98$, $B=10.^{m}08$, $U=10.^{m}10$) were acquired via referencing to photometric standards in NGC~6633 and NGC~7063 (Hiltner et al., 1958). The standard errors in the photometric magnitudes are $\sigma_V=0^{m}.009$, $\sigma_B=0^{m}.009$, $\sigma_U=0^{m}.012$. The measurements were carried out with an aperture of $27^{\prime\prime}$ (sometimes $13^{\prime\prime}$), so we measured the integral brightness of the object -- the PN and the central star.  

Here we present the magnitudes in the instrumental system which is very close to the standard one of Johnson. Standard data reduction procedures were performed, and besides, the magnitudes were reduced to the common under-dome temperature ($t=+10^{\circ}$C) and a correction was introduced after a slight change of instrumental system in 1989 when the photomultiplier was substituted.

The magnitude -- temperature relations for $B$ and $V$ can be expressed by the equations:

\begin{equation}
\begin{array}{c}
\Delta V=0.121-0.013t+1.624\times10^{-4}t^2,

\Delta B=-0.011-0.001t,

\end{array}
\end{equation}

where $t$ is the under-dome temperature in Celsius degree.

To reduce new data to the previous photometric system we introduce the following corrections: $\Delta V=0.^{m}109$, $\Delta B=-0.^{m}084$,
$\Delta U=-0.^{m}040$.

In Table~\ref{ubv} we present $UBV$ photometry for IC~4997 obtained in 2009-2019. To analyze its long-term variability we calculated the annual average $UBV$ magnitudes and present them in Table~\ref{mean} together with standard deviation ($\sigma$) and number of observations ($N$). 

\subsection*{IR photometry}

We carried out IR photometric monitoring at the 1.25-m telescope of CAS in 1999-2006 and resumed in 2019. The photometer with a liquid nitrogen cooled
photovoltaic indium antimonide (InSn) detector installed at the Cassegrain focus was used. The output aperture was $\approx12^{\prime\prime}$, the spatial separation of the beams during modulation was $\approx 30^{\prime\prime}$ in the east–west direction. The star BS~7635 from Johnson et al. (1966) was used as a photometric standard. The results of IR monitoring in 1999-2006 were published in Taranova and Shenavrin (2007). New $JHKL$ magnitudes obtained in 2019 are listed in Table~\ref{IR}.

\subsection*{Spectral observations}

Optical spectroscopy of IC~4997 was carried out at the 1.25-m telescope of CAS in 2010-2019. We used a low-dispersion spectrograph with a 600 lines mm$^{-1}$ grating. The detector was a ST-402 CCD 765$\times$510 pixels in size. The slit width was constant and equal to $4^{\prime\prime}$. The spectral resolution obtained was near 7.4~\AA\ as measured from the fullwidth half maximum (FWHM). The spectrograph configuration was different in August and October, 2019: the spectra were obtained with another camera objective lens and a FLI PL-4022 CCD 2048$\times$2048 pixels in size. A binning of 2$\times$2 was used, obtaining nearly the similar spectral resolution as before. For flux calibration spectrophotometric standard stars with known spectral energy distribution were also observed: 107~Her, 18~Vul, 29~Vul, HD~196775, 40~Cyg, $\rho$~Aql (Glushneva et al., 1998; Pickles, 1998). The standard stars were observed before or after IC~4997 at close airmasses. We show the log of observations in Table~\ref{sp_log}. 

Due to the spectrograph design it's possible to obtain simultaneously a part of spectrum of $\sim1500$~\AA\ or $\sim2400$~\AA\ in the previous or new configuration respectively. To cover the whole observable wavelength range ($\sim$4000--9500~\AA) it's necessary to obtain 4 or 3 overlapping frames. Usually the airmass difference between IC~4997 and the standard star did not exceed 0.2. Under stable  atmospheric conditions the estimated error in the absolute flux calibration was about $\sim$5\%, as is was indicated by good matching of overlapping parts of spectra. Under non-stable conditions the calibration error increased (up to 20\%).   

To measure the emission line intensities we integrated all the flux in the line. Table~\ref{intensity} presents observed relative emission line intensities on the scale  $I(\textrm{H}\beta)=100$, and the observed flux $F(\textrm{H}\beta)$ in absolute units. For the lines brighter than 1\% of the H$\beta$ line the estimated error is about 10\%, and about 25\% for the weaker ones.   

In order to examine the physical conditions in the nebula, we had to correct the relative intensities for the interstellar reddening. Burlak and Esipov (2010) reviewed the previous attempts to determine the reddening coefficient for IC~4997, and using their spectral data obtained in 2003--2009 they derived $c(\textrm{H}\beta)=0.35$ taking into account self-absorption in the hydrogen lines. Our new spectral data agree well enough with this value of $c(\textrm{H}\beta)$. It's worth mentioning that the Balmer decrement was indicative of self-absorption in 2010--2019, too, but the effect seems to have weakened by 2019. 

\section*{PHOTOMETRIC VARIABILITY}

\begin{figure}
\begin{center}
 \includegraphics[scale=2.2]{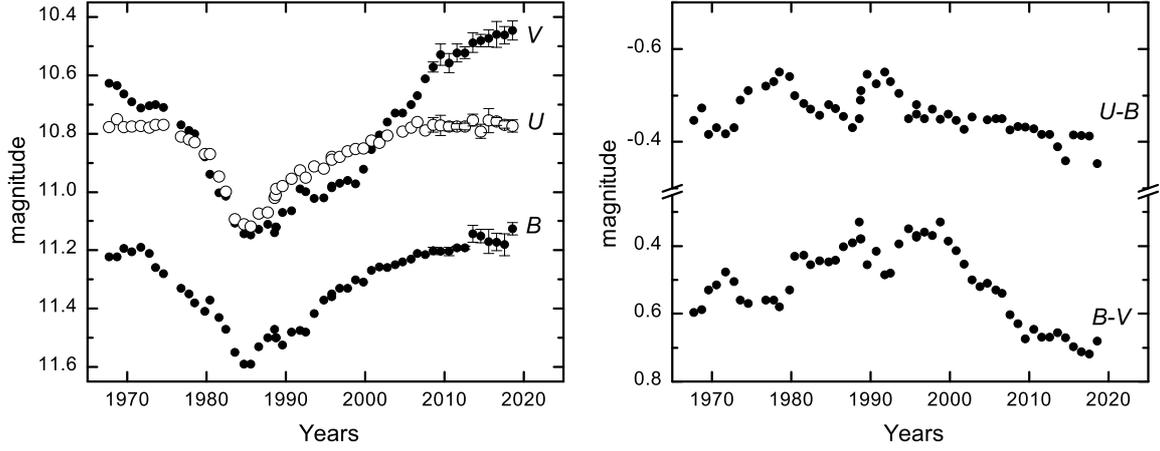}
 \caption{The annual average $UBV$ light and $U-B, B-V$ color curves for IC~4997 in 1968--2019.}
 \label{meanubv}
\end{center}
\end{figure}

The integral brightness of IC~4997 varies with a typical seasonal range of less than 0.$^{m}$2 and demonstrates a long-term trend over the last 50 years. Figure~\ref{meanubv} shows the evolution of the annual average $UBV$ brightness, and $U-B$, $B-V$ color indices in 1968--2019 basing on the results of Kostyakova et al. (1973), Kostyakova (1991, 1999), Arkhipova et al. (1994), Kostyakova and Arkhipova (2009), unpublished measurements of E.B.~Kostyakova and new data.  

The variation of the annual average magnitude appears to have an amplitude of $0.^{m}6$ in $V$, $0.^{m}4$ in $B$, and about $0.^{m}3$ in $U$. Note that in 50 years of our monitoring the integral $U$ and $B$ magnitudes described a gradual asymmetric curve and returned to the initial state, while the $V$ brightness continued to increase after 2010, and the object is $0.^{m}2$ brighter at present than it was when we started observing it. The $B-V$ and $U-B$ color indices varied less gradually and appeared bluer and with larger scatter when the object was fainter that could be ascribed to the effect of variable emission lines. 

To estimate the input from emission into the $UBV$ brightness we isolated the emission nebular component of IC~4997 from the average $UBV$ magnitudes for the year 1990 using the most complete and reliable spectral data for this year from Hyung et al. (1994). We took into account the lines not fainter than 0.05 on the scale $I(\mathrm{H}\beta) = 100$. Unfortunately, the given paper contains only relative line intensities for IC~4997 in August, 1990. The data were flux-calibrated using the absolute measurement of H$\beta$ intensity in 1990 made by Kostyakova and Arkhipova (2009): $F(\mathrm{H}\beta)=2.6\times10^{-11}$erg~cm$^{-2}$~s$^{-1}$. We calculated the averaged values of brightness and color over 12 estimates obtained in May-October, 1990: $V=11.^{m}07$, $B=11.^{m}52$, $U=10.^{m}98$, $B-V=+0.45$, $U-B=-0.54$. The input from emission lines was estimated using the passbands of our $UBV$ instrumental system.  

Note that photometric and spectrophotometric observations were carried out at different nights, the averaged date of all measurements falls on the end of July, 1990.

So, we estimated the emission lines input into the $UBV$ brightness, and calculated the integral continuum flux and magnitude for the object in 1990.  

$F_V(\mathrm{lines})=4.57\times10^{-11}$erg~cm$^{-2}$~s$^{-1}$, 
$F_V(\mathrm{cont})= 8.91\times10^{-11}$erg~cm$^{-2}$~s$^{-1}$,  $V(\mathrm{cont})=11.^{m}51$;

$F_B(\mathrm{lines})=11.06\times10^{-11}$erg~cm$^{-2}$~s$^{-1}$, $F_B(\mathrm{cont})=4.73\times10^{-11}$~erg~cm$^{-2}$~s$^{-1}$, $B(\mathrm{cont})=12.^{m}83$;

$F_U(\mathrm{lines})= 2.35\times10^{-11}$erg~cm$^{-2}$~s$^{-1}$, $F_U(\mathrm{cont})=17.11\times10^{-11}$erg~cm$^{-2}$~s$^{-1}$, $U(\mathrm{cont})=11.^{m}14$.

The emission lines input into the $UBV$ integral brightness of the variable PN IC~4997 in 1990 was as large as 30~\% in $V$, 70~\% in $B$, and more than 13~\% in $U$ (since no emission lines were measured beyond the Balmer discontinuity in 1990).

\begin{figure}
\begin{center}
 \includegraphics[scale=1.2]{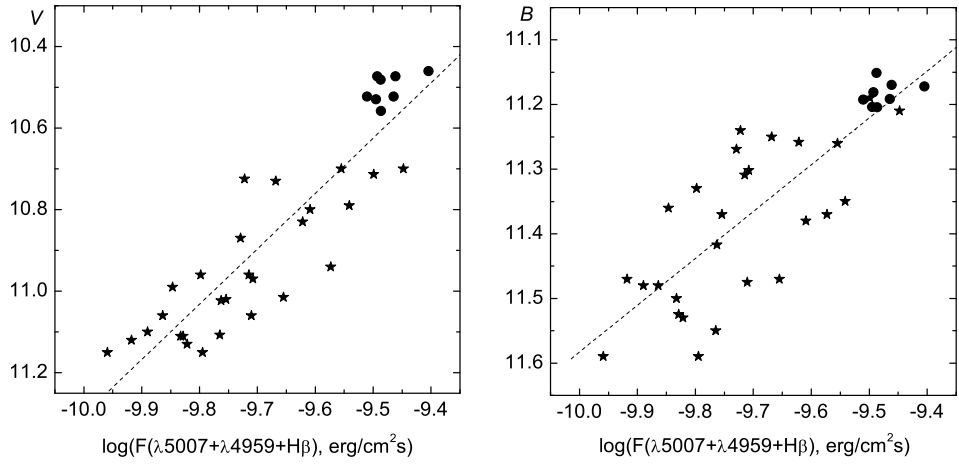}
 \caption{The relationship between $\lg(F(H\beta)+F(\lambda4959)+F(\lambda5007)$) and the $V$ and $B$ brightness in 1972--2019. The asterisks represent the data from Kostyakova and Arkhipova (2009), the points -- this work.}
 \label{corr}
\end{center}
\end{figure}

We'd like to highlight the input of the [OIII] nebular lines $\lambda$5007, $\lambda$4959 and H$\beta$ into the $B$ and $V$ integral light. In Figure~\ref{corr}, we present the relationship between the summary flux from H$\beta$, [OIII] $\lambda$4959, $\lambda$5007, which dominate the blue region, and the annual average $V$ and $B$ brightness in 1972-2019. It is clearly seen that the summary flux from these three lines $\lg(F(H\beta)+F(\lambda4959)+F(\lambda5007))$ and the $V$ and $B$ brightness are closely related: the correlation coefficients are 0.90 and 0.81 respectively. Therefore, the $B$ and $V$ photometric variability may be explained by the changing emission spectrum of the nebula. Similar analysis for the $U$ band is complicated by the absence of data on the changes of nebular spectrum in the wavelength range bluer than $\lambda$3700.      

The color indices for the summary continuum of the nebula and central star appeared to be $B-V=+1.^{m}32$, $U-B <–1.^{m}69$. Assuming $E(B-V)=0.^{m}24$ for IC~4997, we got the reddening corrected color indices for the summary continuum: $(B-V)_0=+1.^{m}08$, $(U-B)_0<–1.^{m}89$. 

The emission lines input having been removed, the position of IC~4997 in the color-color diagram appears to be similar to that of a hot star with a temperature more than 35000~K and a rather optically thick gas continuum. We estimate the position uncertainty associated with the mismatching of dates of photometric and spectral observations to be equal to $\sim 0.^{m}3$.  

Note that IC~4997 was undoubtedly classified as a bipolar PN (Sahai et al., 2011). The bipolar PNs which have an hourglass-like morphology (i.e. a cylinder with a 'pinched-in' shape in the region around the center) are now widely considered to have a binary central star. And in this regard, we were interested to find some traces of the possible central star binarity in the summary optical continuum. The color indices $(B-V)_0(\mathrm{cont})= 1.08$ and $(U-B)_0(\mathrm{cont})<-1.89$ resulted from subtracting the emission lines input and reddenning correction demonstrate a slight red excess in the $(B-V)$ color which is not reliable due to low accuracy.       

\begin{figure}
\begin{center}
 \includegraphics[scale=1.5]{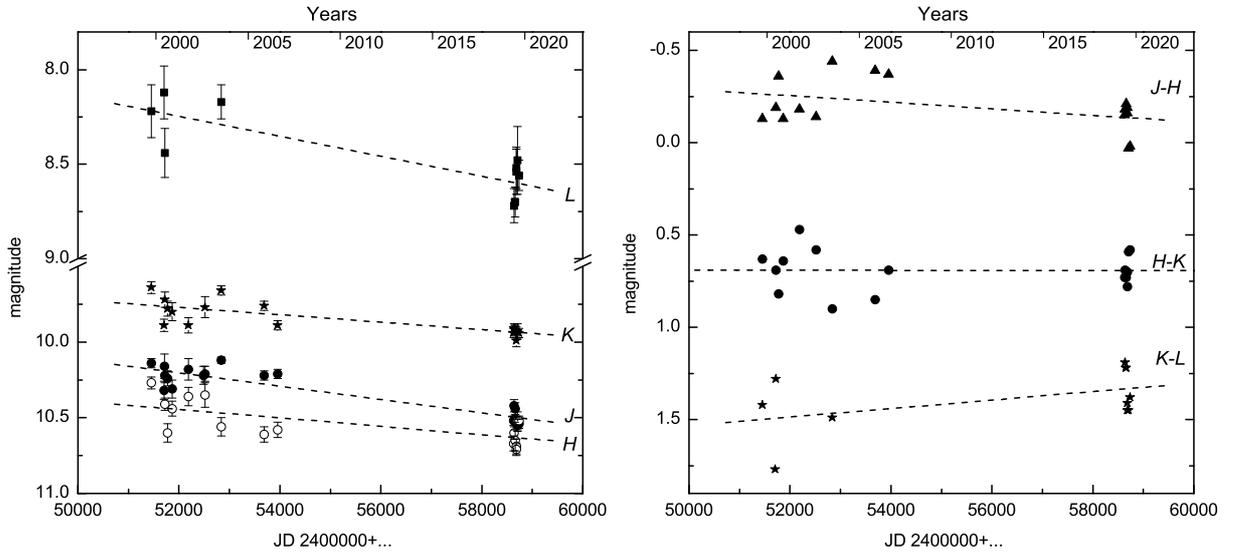}
 \caption{The change in near IR brightness and color for IC~4997 in 1999--2019.}
 \label{nir}
\end{center}
\end{figure}

IC~4997 displays a variation in IR brightness, too. Taranova and Shenavrin (2007) basing on the $JHKL$ photometry made in 1999-2006 showed that the amplitude of variation in $JHKL$ was nearly $0.^{m}2-0.^{m}3$ and that in the $J-H$, $H-K$ и $K-L$ colors was up to $\sim 0.^{m}5$ within a characteristic time of 260--280 days. The authors suggested that the detected variation in $H$ was associated with the changing input of the hydrogen Brackett lines. In 2019 we carried out the observations of IC~4997 with the same equipment as had been used previously. In Figure~\ref{nir} we present the near IR light and color curves for IC~4997 compiled from the data published by Taranova and Shenavrin (2007) and the new measurements obtained in 2019. As is seen from the figure, the $H$ and $K$ brightness in 2019 is at the level of the minimum values for the 1999-2006 interval, whereas the average $J$ brightness has decreased by $0.^{m}25$. The most prominent change is seen for the $L$ band: the brightness has decreased by $0.^{m}4$. The color indices have barely changed on average since 1999 and are about $J-H=-0.^{m}2$, $J-K=0.^{m}5$, $H-K=0.^{m}7$, only the $K-L$ color has decreased from $1.^{m}5$ to $1.^{m}3$.

Whitelock (1985) reviewed the principal sources of near IR emission for PNs: they are the free-free and free-bound radiation of hydrogen and helium plasma and the thermal emission from dust with $T_{d}$>1000~K (if there is dust in the nebula). There is also a small input of the central star and a significant one from the emission lines, the strongest ones being:

$J$ band: P$\beta$, P$\gamma$, He~II $\lambda1.162\mu$m, He~I $\lambda1.083\mu$m;

$H$ band: Brackett series from Br~10 $\lambda1.737\mu$m to nearly the series limit at $\lambda1.459\mu$m;

$K$ band: Br$\gamma$ and He~I $\lambda2.058\mu$m.

Whitelock (1985) proposed a classification scheme for PNs based on the major source of 1-2$\mu$m emission and assigned IC~4997 to the N (Nebula) group due to the nebula radiation dominating the near IR domain. Note that IC~4997 is a PN of low excitation, so the He~II $\lambda1.162\mu$m is not present in the spectrum. Thus, trying to explain the $JHK$ variability of IC~4997 we can ignore the impact of the dust component and consider the change in intensities of H~I and, to a lesser degree, of He~I lines responsible for variation.

Ohsawa et al. (2016) presented near IR ($2.5-5.0\mu$m) spectra for 72 PNs including IC~4997 obtained with IRC/AKARI. Due to this work, we have an idea of the most important emission lines of IC~4997 falling into the $L$ passband: they are Br$\alpha$ $\lambda4.05\mu$m and Br$\beta$ $\lambda2.62\mu$m. Therefore, the $L$ brightness variability of IC~4997 may relate to the change in the intensity of these lines.   

\section*{SPECTRAL VARIABILITY}

\begin{figure}
\begin{center}
 \includegraphics[scale=1.2]{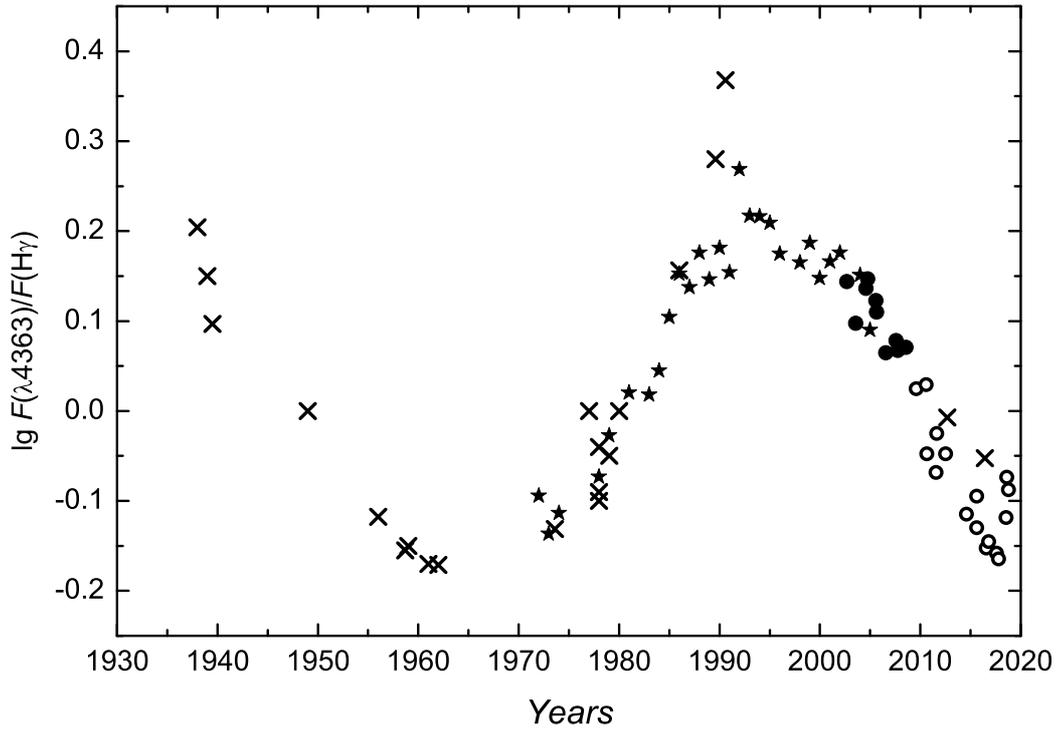}
 \caption{The evolution of the observed intensity ratio  $F(\lambda4363)/F(\textrm{H}\gamma$) based on data from different studies: asterisks -- Kostyakova, Arkhipova (2009), filled circles -- Burlak, Esipov (2010), open circles -- this work, crosses -- other sources, see the references in the text.}\label{lg_r}
\end{center}
\end{figure}

The change in the intensity ratio of the lines [OIII] $\lambda$4363 and H$\gamma$ was the first evidence for the spectral variability of IC~4997 due to the fact that it could be easily measured with confidence. In Figure~\ref{lg_r} we reconstruct the long-term (several decades) evolution of the $F(\lambda4363)/F(\textrm{H}\gamma)$ ratio basing on the data from this work and the previous estimates made by different researchers since 1938: Aller (1941), Struve and Swings (1941), Page (1942), White (1952), Liller and Aller (1963), Aller and Liller (1966), Vorontsov-Velyaminov et al. (1965), Aller and Kaler (1964), O'Dell (1963), Ahern (1978), Feibelman et al. (1979), Purgathofer (1978), Purgathofer and Stoll (1981), Acker et al. (1989), Hyung et al. (1994), Kostyakova and Arkhipova (2009). Besides, we have estimated line intensities in the spectrum of IC~4997 published by Hajduk et al. (2015) and in another one found in the HASH PN database (Parker et al., 2016). Over the entire period of observations the ratio has described a wave with a peak-to-peak amplitude of 0.45 in the logarithmic scale and a characteristic time of 50-60 years. The $\lambda4363/\textrm{H}\gamma$ ratio continued to decrease in 2010-2019 following the trend set around 1990, and by 2019 the ratio has reached the value observed in 1960--1970. 

\begin{figure}
\begin{center}
 \includegraphics[scale=1.1]{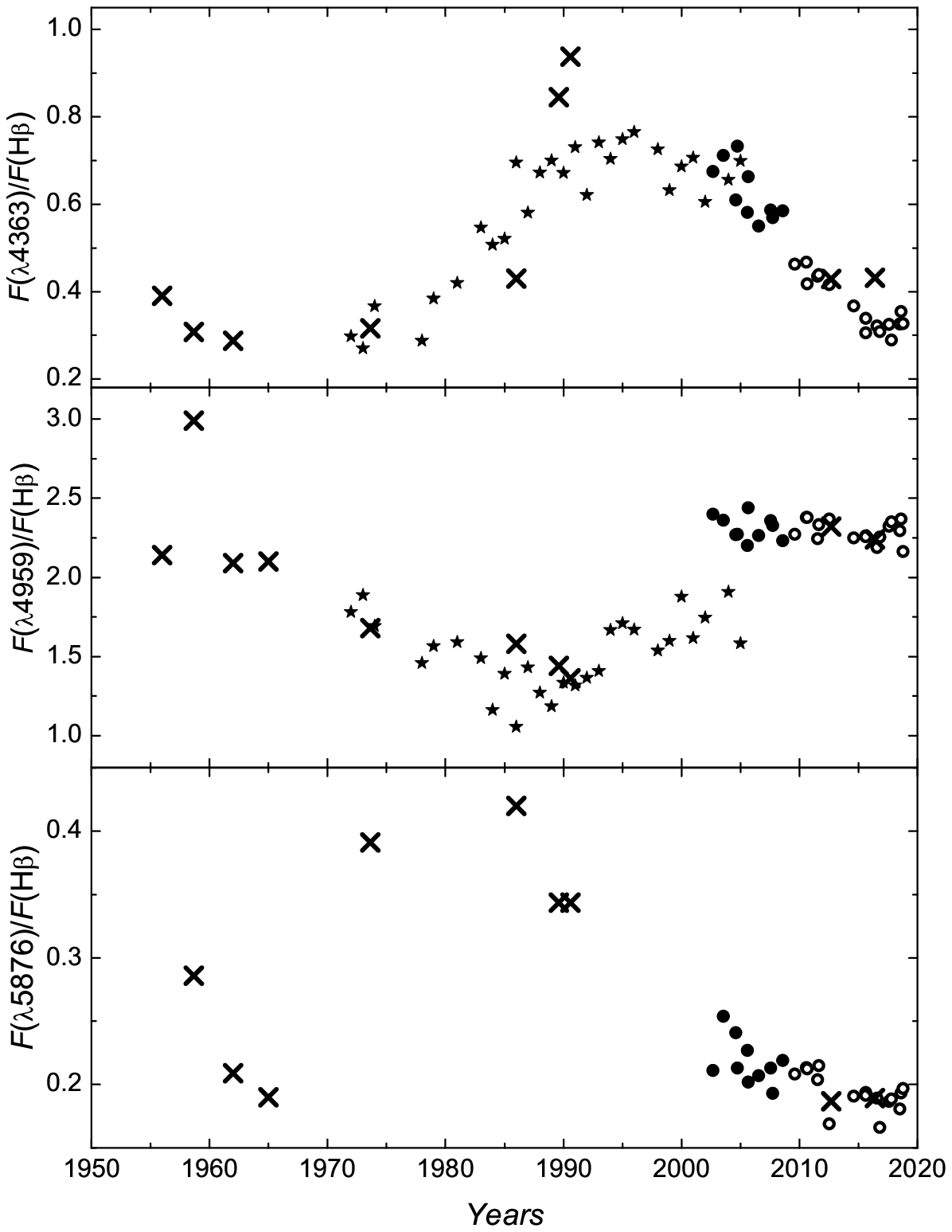}
 \caption{The evolution of the observed relative intensities of [OIII] $\lambda4363$, $\lambda4959$ and HeI $\lambda5876$ based on data from various sources. The symbols are the same as for Figure~\ref{lg_r}.}
 \label{rel}
\end{center}
\end{figure}

In addition to the variation of $F(\lambda4363)/F(\textrm{H}\gamma)$, the nebular [OIII] lines intensities relative to H$\beta$ were found to vary, too. Since 1960--1970 their intensities at first decreased until 1985, then they grew to 2005, and since then they stay at the approximately constant level slightly higher than that of the early sixties. Note that the relative intensities of the nebular and auroral lines have changed differently which is well illustrated by Figure~\ref{rel}. There is also a hint of variability in He~I lines but the data are less reliable and available for a smaller time interval. In Figure~\ref{rel} we present the evolution of the He~I $\lambda5876$ intensity relative to H$\beta$: in general, the change was inverse to that of the nebular lines (Vorontsov-Velyaminov et al., 1965; O'Dell, 1963; Aller and Walker, 1970; Ahern, 1978; Acker et al., 1989; Hyung et al., 1994; Hajduk et al., 2015; Parker et al., 2016). The He~I $\lambda6678$ line varies in a similar way to $\lambda5876$, at least since 1986 -- we have not managed to find earlier measurements for the $\lambda6678$ line.    

\begin{figure}
\begin{center}
 \includegraphics[scale=1.1]{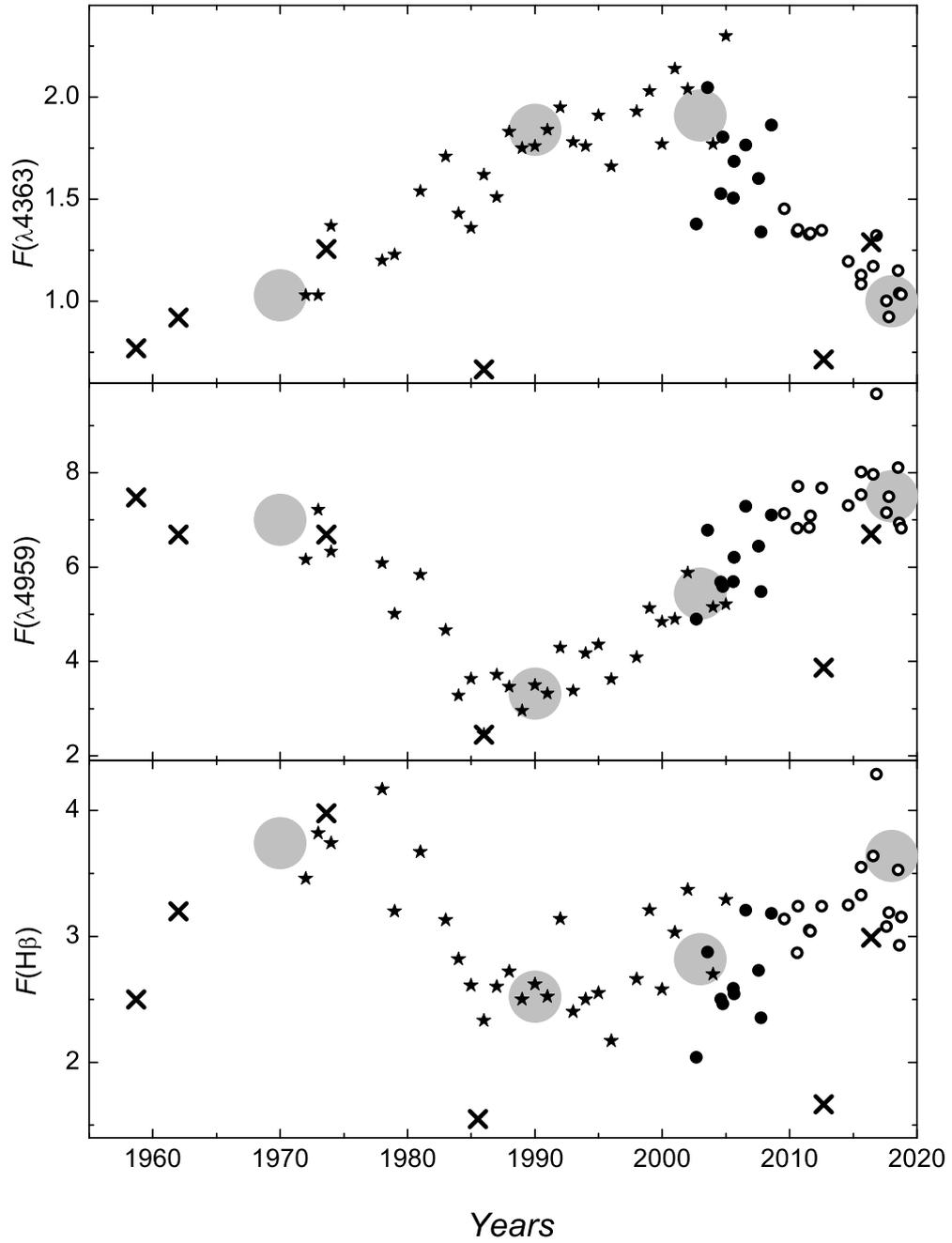}
 \caption{The evolution of the observed absolute intensities of [OIII] $\lambda4363$, $\lambda4959$ and H$\beta$ expressed in units of $10^{-11}$~erg$~$cm$^{-2}$s$^{-1}$. The symbols are the same as for Figure~\ref{lg_r}. Big grey circles represent average values at the points that delimit the time intervals distinguished by the peculiarities of [OIII] $\lambda4363$ and $\lambda4959$ variations.}
 \label{abs}
\end{center}
\end{figure}

Figure~\ref{abs} shows the observed absolute intensities for the [OIII] $\lambda4363$, $\lambda4959$ lines measured by different researchers over the last 60 years. Although there is a big scatter in estimates for these lines, the trend is well defined. Since 1960 the absolute intensity of the auroral line grew at an approximately constant rate, reached maximum level about 2000--2005 having increased more than twice, and then started to decrease faster than it had been increasing. Since 1960 the absolute intensities of nebular lines stayed quite constant with a possible slight tendency to decrease, but in the mid-1970s dropped sharply and reached minimum level around 1985-1990, then increased at a lower rate to the values observed about 1960 or little higher, thereby having reproduced the optical light curve of IC~4997.   

It would be interesting to ascertain whether the absolute intensity of $\textrm{H}\beta$ varies, but the absolute spectrophotometry is very difficult to perform, and there are inconsistencies between observers which produce large scatter in data constituting definite obstruction to the determination of the shape and characteristic parameters of the $\textrm{H}\beta$ intensity variation. Nevertheless, one can see a period of higher absolute intensity before 1980, then it faded by 30\% and stayed more or less constant until 2000--2005, and after that started to grow.     

\begin{figure}
\begin{center}
 \includegraphics[scale=1.1]{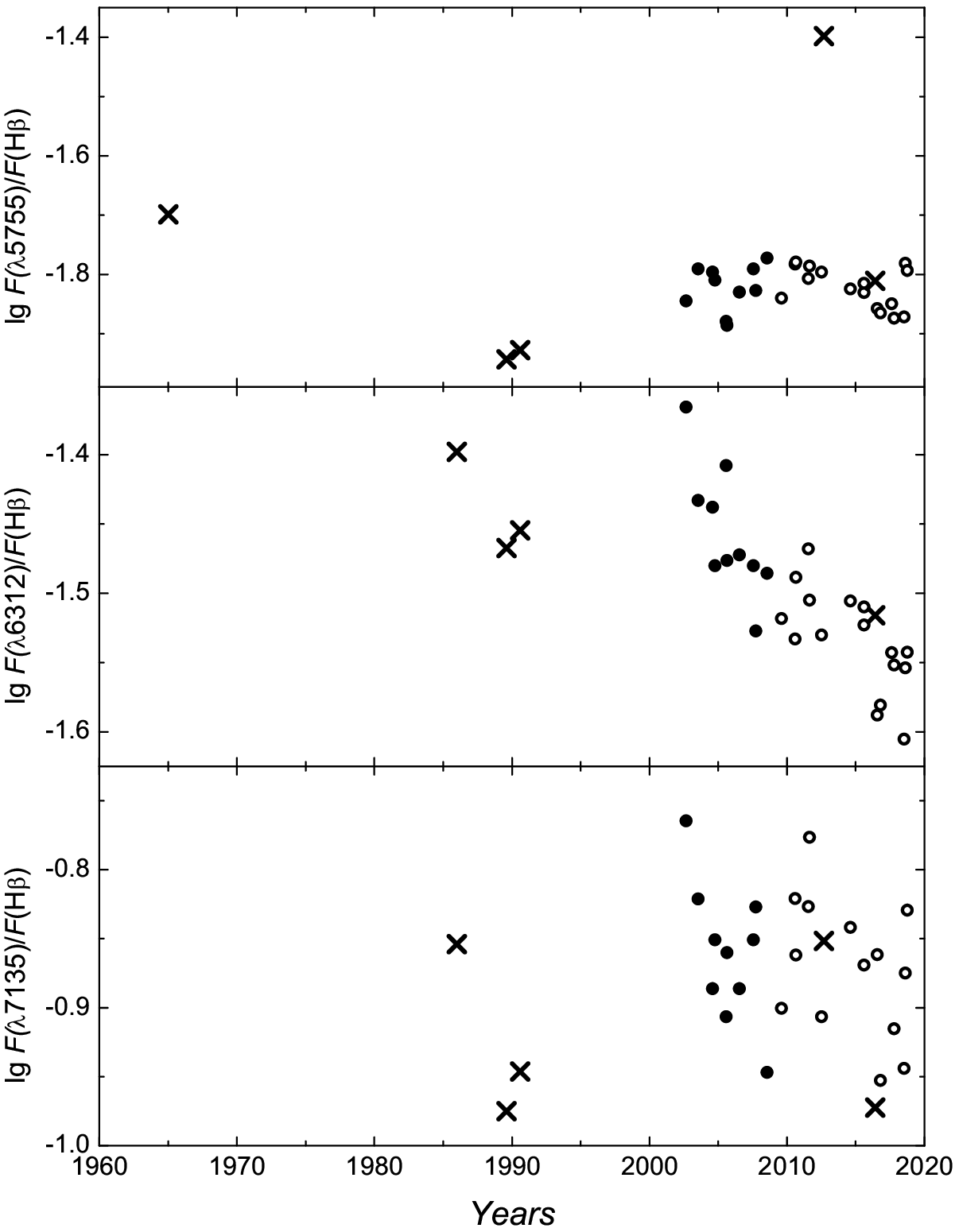}
 \caption{The evolution of the observed relative intensities of [ArIII] $\lambda7135$,  [SIII] $\lambda6312$, [NII] $\lambda5755$ based on data from various sources. The symbols are the same as for Figure~\ref{lg_r}.}
 \label{Ar_S_N}
\end{center}
\end{figure}

As regards the variability of other lines, it's worth mentioning that mostly short wave region of optical spectrum was investigated in earlier epochs, while the long wave one -- later. So, just several lines were measured over a long-term period, and only for a few the estimates are reliable. Figure~\ref{Ar_S_N} shows the evolution of relative intensities for the [SIII] $\lambda6312$ and [NII] $\lambda5755$ auroral lines, and the [ArIII] $\lambda7135$ nebular line. In 1986--2019 [SIII] $\lambda6312$ reproduced the behavior of the [OIII] auroral line, its relative intensity decreased, whereas the variation of [NII] $\lambda5755$ was inverse to that of [OIII] $\lambda4363$. We have not managed to ascertain the change in [ArIII] $\lambda7135$ relative intensity over the given time interval. 

\section*{DIAGNOSTICS AND PHYSICAL PARAMETERS OF THE IC~4997 NEBULA}

       \begin{flushright}
"One has to be an optimist to attempt to construct \\a model for IC~4997..." \\
                    (Huyng, Aller, Feibelman, 1994)
        \end{flushright}

After the variability of the $F(\lambda4363)/F(\rm H\gamma)$ and $F(\lambda4363)/F(\lambda4959+\lambda5007)$ ratios was discovered and till about 1970, some suggestions were proposed to explain these spectral changes. Gurzadian (1958), Vorontsov-Velyaminov (1960) and Khromov (1962) considered the fall in the central star temperature to be the cause of the decrease in $F(\lambda4363)/F(\textrm{H}\gamma)$, though Gurzadian believed this change in temperature to be of an evolutionary character, Vorontsov-Velyaminov -- to have a long periodic nature, Khromov -- to be oscillatory fluctuations. According to Aller and Liller (1966), the spectral changes resulted from the expansion of the nebula with a corresponding decrease both in electron density and electron temperature. In addition to long-term changes, fast oscillations of the $F(\lambda4363)/F(\textrm{H}\gamma)$ ratio within one year were detected, and Ferland (1979) supposed that they were due to the variation of electron temperature attributed to small changes in the number of ionizing photons emitted by the central star: since the electron density is high in the inner parts of the nebula ($N_e\sim 10^6$~см$^{-3}$) where the [OIII] $\lambda$4363 line is mostly generated, the thermal equilibrium is set very quickly and a well-defined temperature will always be present.          
Although there is a variety of diagnostic ratios measured for IC~4997 in the optical and ultraviolet regions, the constructing of a quantitative model turned out to be a difficult task, especially if one aims not only to describe the instantaneous state of the nebula and central star but also to interpret its change (see Hyung et al., 1994 for more detail). Various ions emit in different zones and there exists a large range in electron density and temperature for them. So, the region of intersection of all the curves corresponding to different ions encompasses a large range of $N_e, T_e$ values in the diagnostic diagram for IC~4997. To understand the observational data and to estimate the abundances it is necessary to adopt for the nebula at least a two-component structure consisting of a more dense inner zone enclosed in a shell of lower density.

\begin{figure}
\begin{center}
 \includegraphics[scale=1.2]{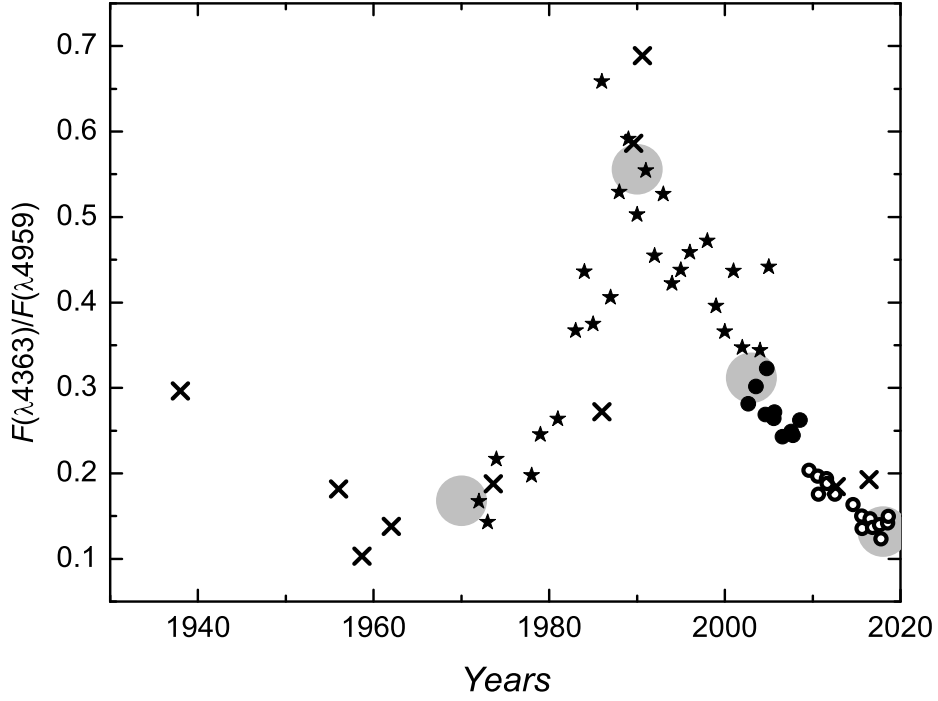}
 \caption{The evolution of the observed auroral to nebular line ratio $F(\lambda4363)/F(\lambda4959)$ based on data from various sources. The symbols are the same as for Figures~\ref{lg_r} and \ref{abs}.}
 \label{Aur_N2}
\end{center}
\end{figure}

Our aim was to determine the physical conditions in the nebula, so we measured several diagnostic ratios. The most actively used intensity ratio has always been the relation involving the auroral [OIII] $\lambda4363$ and one or both of the nebular [OIII] $\lambda4959, 5007$ transitions. We did not always manage to measure the $\lambda5007$ line intensity, therefore Figure~\ref{Aur_N2} shows the change of the $R=F(\lambda4363)/F(\lambda4959)$ value over a time period of about 80 years. In 2010--2019 the ratio decreased thus keeping the tendency started about 1990, and by 2019 it returned to the value observed in 1960-1970. Due to the availability of absolute intensities, we can schematically divide the evolution of $R$ in 1970-2019 into three intervals. During the first phase ($\sim$1970-1990) $R$ was growing due to the strengthening of the auroral transition and weakening of the nebular one; at the second stage ($\sim$1990-2000) -- $R$ was decreasing with the simultaneous strengthening of both lines; during the third phase ($\sim$2000-2019) -- $R$ was decreasing due to the strengthening of the nebular transition and weakening of the auroral one. The big grey circles in Figures~\ref{abs} и \ref{Aur_N2} indicate the averaged values of corresponding quantities at the points which delimit the highlighted intervals: 1970, 1990, 2003 (the year when we started our spectroscopic monitoring with a CCD), 2019.           

\begin{figure}
\begin{center}
\includegraphics[scale=1.1]{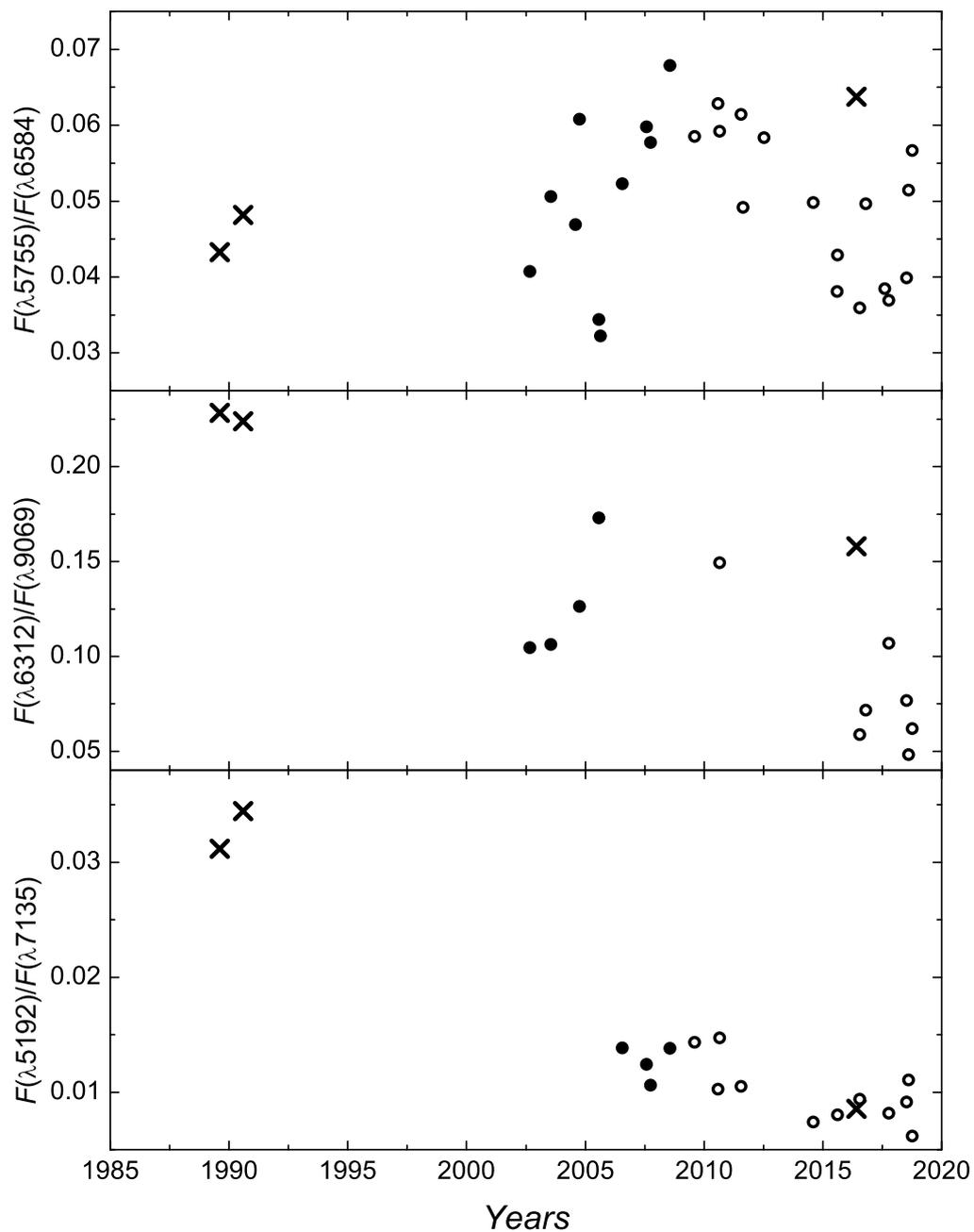}
\caption{The evolution of the observed auroral to nebular line ratios [ArIII] $F(\lambda5192)/F(\lambda7135)$, [SIII] $F(\lambda6312)/F(\lambda9069)$, [NII] $F(\lambda5755)/F(\lambda6584)$ based on data from various sources. The symbols are the same as for Figure~\ref{lg_r}.}
 \label{R_ArSN}
\end{center}
\end{figure}

Some other line ratios seem to be variable too, though their variation can be traced over a shorter period of time and is not evident (see Figure~\ref{R_ArSN}). So, the [ArIII] $F(\lambda5192)/F(\lambda7135)$ ratio was observed to decrease from $\sim0.033$ in the early 1990s (Hyung et al.,1994) to $\leq0.01$ (this work) in 2019. The [ArIII] $\lambda5192$ line, however, is very weak and forms a blend with two other lines of comparable intensity, [NI] $\lambda5198$ и $\lambda5200$, and the low dispersion makes its measuring uncertain. Similar behavior is seen for the [SIII] lines, though a large wavelength separation between them ($\lambda6312$ and $\lambda9069$) leads to a significant calibration uncertainty. On the contrary, the [NII] $F(\lambda5755)/F(\lambda6584)$ ratio does not show any significant change. Besides, the scatter of estimates is big, possibly due to the difficulty in separating the $\lambda6584$ line from the H$\alpha$ wing.        
       
{\sloppy The diagnostic relations sensitive to electron density, [SII] $F(\lambda6716)/F(\lambda6731)$ and [ClIII] $F(\lambda5518)/F(\lambda5538)$, stayed constant within the limits of accuracy over the whole period of our observations since 2003. The $F(\lambda6716)/F(\lambda6731)$ ratio is close to its critical value, so it indicates only the lower limit of $\log N_e\sim4$ which corresponds to the outer envelope of lower density where the low excitation species emit. The ratio of [ClIII] lines is also close to its critical value. This ratio is attributed to the zones of higher excitation and suggests a somewhat higher estimate for the density of $\lg N_e\sim4.5$. But the second ionization potential for chlorine is lower than that for oxygen, therefore the zone where the $\lambda5518,5538$ lines originate may match only partially with the [OIII] emitting region. The constancy of the ratios mentioned above implies that even if the electron density in the low excitation zones varies it does not fall much lower than the critical value ($\lg N_e^{crit}\sim3-3.5$ for sulphur и $\sim 4-4.5$ for chlorine).          

}

\begin{figure}
\begin{center}
 \includegraphics[scale=0.8]{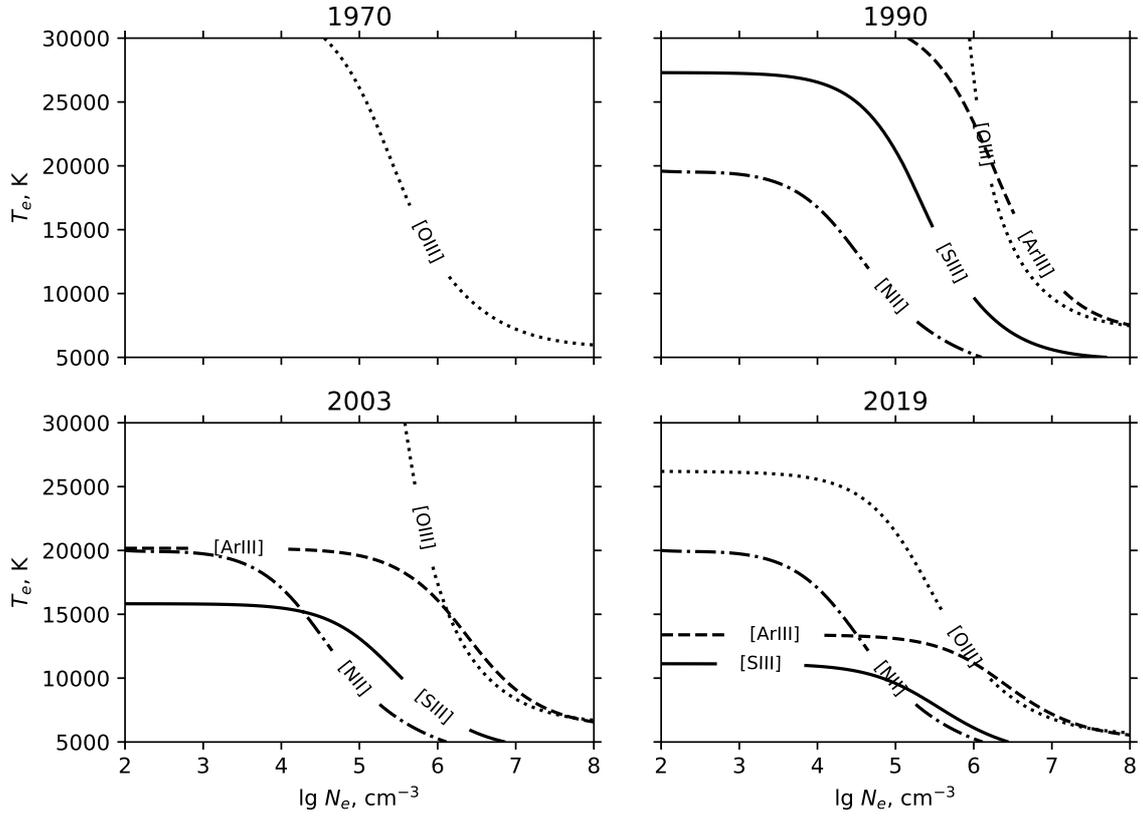}
 \caption{Diagnostic diagrams for IC~4997 drawn with the use of averaged data from different sources for the epochs of 1970, 1990, 2003 and 2019. All the intensity ratios were corrected for interstellar reddening with $c(\textrm{H}\beta)=0.35$. Different lines indicate ions: solid for [SIII], dotted for [OIII], dashed for [ArIII], and dotted-dashed for [NII].}
 \label{diag}
\end{center}
\end{figure}

Let's now consider the displacement of diagnostic curves representing the ions N$^+$, S$^{2+}$, O$^{2+}$ and Ar$^{2+}$ in the $N_e-T_e$ diagram. These ions have a similar structure of energetic levels but differ in the ionization potential and critical density, and must emit in different regions of the nebula. Figure~\ref{diag} contains the $N_e-T_e$ diagrams constructed by means of the PyNeb package (Luridiana et al., 2015) for the moments which limit the intervals of time highlighted according to the change of the [OIII] absolute intensities. Besides the fact that one pair of $N_e, T_e$ values can not explain the observed data, the total appearance of the diagrams suggests a high value of the electron density ($\lg N_e\geq6$), at least for the strata where the [OIII] lines arise. The ratios used for the diagrams were corrected for the interstellar reddening with $c(\textrm{H}\beta)=0.35$. The diagram for the year 1990 was drawn basing on the data from Hyung et al. (1994), and they derived $c(\textrm{H}\beta)=0.8$. If we had corrected the ratios using this value of $c(\textrm{H}\beta)$, the diagnostic curves would have been located in the area of larger values of $N_e, T_e$. Since 1990 the curves associated with O$^{2+}$, S$^{2+}$ и Ar$^{2+}$ evolved in similar manner, they shifted to the smaller values of $N_e, T_e$, while the location of the N$^{+}$ curve almost did not change. Probably, the variation of $N_e, T_e$ does not affect the [NII] emitting zone.              

As we are not able to derive the absolute $N_e, T_e$ values for the given epochs, let's try to determine how they change. The initial data for our further calculations are the changes of absolute intensities for $\textrm{H}\beta$, [OIII] $\lambda4363, 4959$, and also the changes of their relations corrected for reddening. We assume that the zones where all these transitions arise are characterized by the same $N_e, T_e$ values and the number of emitting hydrogen atoms is constant. Some definite value of temperature is considered as an initial one (in 1970). First, we find $N_e$ for this initial moment using $R$, and the relative abundance of $\rm O^{2+}$ using relative intensities. Then, as we know the relative change of $F(\textrm{H}\beta)$ between the two epochs and the $R$ value at the second moment, we can estimate $N_e, T_e$ at the second moment and also the abundance of $\rm O^{2+}$ using relative intensities. The same procedure is executed as we pass from the second moment to the third and from the third to the fourth. Table~\ref{NeTe} lists the results of the simulations performed by means of the package PyNeb for several initial values of the electron temperature: $T_e=8000, 10000, 12000$~K. For all initial temperatures, $N_e$ varies only slightly, whereas $T_e$ first grows and then drops by several thousand degrees.       

A significant growth of temperature is necessary to interpret the decrease in the absolute intensity of $\textrm{H}\beta$ since 1970. It сould have been explained by the decrease in $N_e$, but it is in contrary to the growth of $R$. On the other hand, the intensity of collision excited lines must increase with temperature, but we observe the nebular lines decreasing since 1970, and the auroral line increasing is less than expected for such change in temperature. To explain the observed data it's necessary to reduce significantly the number of emitting O$^{2+}$ ions (by a factor of 10), which may be due to the subsequent ionization of oxygen or to the diminution of the emitting zone. 

And the question arises: what did make $T_e$ to increase by several thousand degrees? According to Ferland (1979), if the change in the electron temperature of the nebula is caused by a change in the flux of ionizing photons $Q$ from the central star, we can relate them through an equation: $\Delta  T_e\simeq4200\times\Delta Q/Q$. So, the 300~K temperature change proposed by Ferland to explain fast variability of the $F(\lambda4363)/F(\textrm{H}\gamma)$ ratio requires an 8\% change in the ionizing flux. But our simulations imply a temperature increase of several thousand degrees that would mean ionizing flux increasing by several times. We find it hard to imagine such a process. A rise in $T_e$ must have been caused not so much by the growth in the effective temperature of the star because the level of excitation of the spectrum of IC~4997 did not change over the observed period, but by some other processes, for instance, the interaction of stellar winds.    

Table~\ref{NeTe} also lists the relative intensities for the HeI $\lambda5876$ line calculated using the obtained values of $N_e, T_e$ and $\rm He^+/\rm H^+$=0.1. One can see that in the case of the initial $T_e=8000$~K the intensity changes only slightly, whereas for $T_e\geq10000$~K the simulated variation of intensity matches qualitatively with the observed one, though the amplitude is smaller.   

The results presented in Table~\ref{NeTe} can be considered as a kind of estimation. A larger value of interstellar reddening ($c(\textrm{H}\beta)>0.35$) will require a larger increase in electron temperature to explain the observed spectral changes. If we take into account the fact that we measure the integral flux from the nebula, whereas the fluctuation of $N_e, T_e$ may occur only in some strata, then the amplitude of temperature change will be larger.   

\section*{DISCUSSION AND CONCLUSIONS}

We have presented optical and near IR photometric and low-resolution spectroscopic data for IC~4997 obtained in 2009--2019. New results are investigated together with the previously published data.  

Basing on the observational data obtained with our invariable $UBV$ photometric system we have plotted an annual average light curve of IC~4997 from 1970 till 2019. We have found a long-term marked dip in $UBV$ flux with a peak-to-peak range of $0.5^{m}$ in $V$. The source started fading after 1970, reached the minimum brightness near 1985, then recovered to the initial level in $B$ and $U$, whereas the $V$ brightness continued growing up to 2019. In 2019 we obtained new near IR $JHKL$ photometric data. The nebula was found to be fainter in 2019 if compared to the epoch of 1999--2006 with the effect most prominent in $L$. We have shown that the long-term optical and IR brightness variation is related mostly to the changing input from the nebular emission lines.  

After the emission contribution was subtracted from the $B$ and $V$ brightness, the $B-V$ color of the integral continuum corrected for reddening appeared still too red for a sum of the stellar and gas continua, which may be indicative of the presence of one more source of continuum radiation in this wavelength range (a satellite of the central star?), although the evidence is very uncertain. It's worth mentioning that the central star of IC~4997 is not only suspected to be binary but it was also included in the list of PN possibly shaped by a triple stellar system (= {\it maybe triple} class) with a probability of 0.33.      

Basing on new and previously published data we have traced the evolution of relative and absolute intensities and their relations for some emission lines originated in the nebula over a period 1970--2019. In 2010--2018 the $\lambda4363/\textrm{H}\gamma$ ratio decreased as it had been doing since about 1990 and by 2019 returned to the value observed in 1960--1970. Over the whole period of observations the ratio drew a wave with a peak-to-peak amplitude of 0.45 in logarithmic scale and a characteristic time of 50-60 years.

With the use of archival and new data we have traced the evolution of absolute intensity for H$\beta$: one can note a period of higher values before 1980, a fading by a factor of 1.5 and subsequent maintenance at the same level with a slight tendency to growth till 2000--2005, then a more pronounced increase. The variation of the absolute intensity of the nebular [OIII] $\lambda4959$ line in 1960-2019 was roughly similar to that of H$\beta$ but had a larger amplitude ($F_{max}/F_{min}\geq2$). Having recovered to the value observed before fading, the absolute intensity of [OIII] $\lambda4959$ did not change significantly in 2010-2019. The absolute intensity of the auroral [OIII] $\lambda4363$ line grew since 1960 at a nearly constant rate, reached maximum value about 2000-2005 having increased more than twice, then started to fade and is still fading.    

We have also reconstructed the variations of intensity for some other transitions. In particular, we have traced the variability of HeI $\lambda$5876 in 1960-2019: an increase in relative intensity by a factor of $\sim$2, then weakening and return to the former level.    

We have estimated the lower limit for the electron density in the outer shell of IC~4997 using the diagnostic ratios of [SII] и [ClIII]: $\lg N_e\sim4$ и $\sim4.5$ respectively. The ratios stayed constant within errors and were close to critical values in 2010-2019.

The location of diagnostic curves of N$^+$, S$^{2+}$, O$^{2+}$ and Ar$^{2+}$ on the $N_e, T_e$ plane indicates that, first, there are zones of different electron temperature and density in the nebula, and, second, the parameters vary with time. A rise in one or both of these characteristics was observed at least in the inner part of the nebula in 1970-1990 with subsequent decrease which lasts until now. Using the data on absolute intensity variations for H$\beta$, [OIII] $\lambda4363, 4959$ and assuming that $N_e, T_e$ are the same for the strata where these lines arise, we have estimated the scale of the variation. We think that the variation of $T_e$ in the inner part of the nebula is responsible for the spectral variability of IC~4997, and the change in $T_e$ is due not so much to the change in ionizing flux from the central star, but to the variable stellar wind and the related processes. In general, the spectral changes observed in 1960-2019 may be interpreted as the observable consequence of a single episode of enhanced mass loss from the variable central star. There are still the questions of what has triggered the increase in mass loss rate, whether this episode was a unique one or could occur again in one form or another.  

\section*{ACKNOWLEDGEMENT}

The authors dedicate this paper to the memory of Doctor of Science Elena~B.~Kostyakova, Senior Researcher Fellow at the Sternberg Astronomical Institute (1924--2013). 

This research has made use of the ADS, SIMBAD and VIZIER databases.
     
\section*{REFERENCES}

\begin{enumerate}

\item A. Acker, J. Koppen, B. Stenholm, and G. Jasniewicz, Astron. Astrophys. Suppl. Ser. {\bf 80}, 201 (1989)

\item A. Acker, J. Marcout,  F. Ochsenbein,  et al., The Strasbourg-ESO Catalogue of Galactic Planetary Nebulae, Parts I, II, (1992)

\item F.A. Ahern, Astrophys. J. {\bf223}, 901 (1978)

\item L.H. Aller, Astrophys. J. {\bf93}, 236 (1941)

\item L.H. Aller and J.B. Kaler, Astrophys. J. {\bf140}, 621 (1964)

\item L.H. Aller and W. Liller, MNRAS {\bf132}, 337 (1966)

\item L.H. Aller and M.F. Walker, Astrophys. J. {\bf 161}, 917 (1970).

\item V.P. Arkhipova, E.B. Kostyakova, R.I. Noskova, Astron. Lett. {\bf20}, 99 (1994)

\item E. Bear, N. Soker, MNRAS {\bf 837}, L10 (2017)

\item M.A. Burlak, V.F. Esipov, Astron. Lett., {\bf 36}, 752 (2010)

\item S. Casassus, L.-Å. Nyman, C. Dickinson, and T. J. Pearson, MNRAS {\bf382}, 1607 (2007)

\item W.A. Feibelman, R.W. Hobbs, C.W. Mc Cracken, L.W. Brown, Astrophys. J. {\bf 231}, 111  (1979)

\item G.J. Ferland, MNRAS {\bf188}, 669 (1979)

\item F.C. Gillett, R.F. Knacke, W.A. Stein, Astrophys. J. {\bf163}, L57-59, 1971

\item I.N. Glushneva, V.T. Doroshenko, T.S. Fetisova, T.S. Khruzina, E.A. Kolotilov, L.V. Mossakovskaya, S.L. Ovchinnikov, and I.B. Voloshina, VizieR Online Data Catalog III/208 (1998).

\item Y. G\'{o}mez, L.F. Miranda, J.M. Torrelles, L.F. Rodr\'{\i}guez, J.A. L\'{o}pez, MNRAS {\bf 336}, 1139 (2002)

\item G.A. Gurzadian G.A., Soviet Astronomy {\bf 2}, 482 (1958)

\item M.Hajduk, P.A.M. van Hoof, and A.A. Zijlstra, Astron. and Astrophys.J.{\bf573} , A65 (2015)

\item W.A. Hiltner,  B. Iriarte, M.L. Johnson, Astrophys. J. {\bf127}, 539 (1958)

\item S. Hyung, L.H. Aller,  Proc. Nat. Acad. Sci. USA {\bf90}, 413 (1993)

\item S. Hyung, L.H. Aller, W.A. Feibelman,  Astrophys. J. Suppl. Ser. {\bf93}, 465 (1994)

\item H.L. Johnson, R.I. Mitchel, B. Iriarte, W.Z. Wisniewski, Comm. Lunar and Planet. Lab. {\bf 4}, 99  (1966).

\item G.S. Khromov, Sov. Astron. {\bf 5}, 619 (1962)

\item E.B. Kostyakova, Sov. Astron. {\bf 14}, 794 (1971)

\item E.B. Kostyakova, Astron. Lett. {\bf 16}, 465 (1990)

\item E.B. Kostyakova, Trudy GAISh, {\bf 62}, 143, (1991)

\item E.B. Kostyakova, Astron. Lett. {\bf 25}, 389 (1999)

\item E.B. Kostyakova, V.P. Arkhipova, Astron. Rep. {\bf 53}, 1155 (2009)

\item Е.B. Kostyakova, V.P. Arkhipova,  M.V. Savel'eva,  Mem. Soc. Roy. Sci. Liege, 6 Ser., {\bf5} , 473 (1973)

\item P. Lenzuni, A. Natta, N. Panagia, Astrophys. J. {\bf345}, 306 (1989)

\item W. Liller and L.H. Aller, Sky and Tel. {\bf16}, 222  (1957)

\item W. Liller and L.H. Aller, Proc. Nat. Acad. Sci. {\bf 49}, 695 (1963)

\item V. Luridiana, C. Morisset, and R.A. Shaw, Astron. Astrophys. {\bf 573}, 42 (2015)

\item V. M. Lyutyi, Soobshch. GAISh {\bf 172}, 30 (1971)

\item W.L.F. Marcolino, and F.X. de Ara\'{u}jo, Astron. J. {\bf126}, 887 (2003)

\item W.L.F. Marcolino, F.X. de Ara\'{u}jo, H.B.M. Junior, and E.S. Duarte, Astron. J., {\bf134}, 1380 (2007)

\item D.H. Menzel, L.H. Aller, M.H. Hebb,  Astrophys. J. {\bf93}, 230 (1941)

\item L.F. Miranda, J.M. Torrelles, C. Eiroa, Astrophys. J. {\bf461}, L111 (1996)

\item L.F. Miranda, J.M. Torrelles,  Astrophys. J. {\bf496}, 274 (1998)

\item A. Natta, N. Panagia, Astrophys. J. {\bf248}, 185 (1981)

\item C.R. O’Dell, Astrophys. J. {\bf 138}, 1018 (1963)

\item R. Ohsawa, T. Onaka, I. Sakon, M. Matsuura, H. Kaneda, Astron. J. {\bf 151}, 93 (2016)

\item T.L. Page, Astrophys. J. {\bf 96}, 78 (1942)

\item Q.A. Parker, I.S. Boji\v{c}i\'{c}, D.J. Frew, Journal of Physics: Conference Series {\bf 728}, article id. 032008 (2016)

\item B.M. Pazderska, M.P. Gawro\'{n}ski, R. Feiler, M. Birkinshaw, I.W.A. Browne, R. Davis, A.J. Kus, K. Lancaster, S.R. Lowe, E. Pazderski, M. Peel and P.N. Wilkinson, Astron. Astrophys. {\bf 498}, 463 (2009) 

\item A.J. Pickles, Publ. Astron. Soc. Pacific {\bf110}, 863 (1998).

\item S.R. Pottasch, B. Baud, D. Beinteme, J. Emerson et al., Astron. Astrophys. {\bf138}, 10 (1984)

\item A.T. Purgathofer, Circ. IAU {\bf3258} (1978)

\item A.T. Purgathofer and M. Stoll, Astron. Astrophys. {\bf99}, 218 (1981)

\item R. Sahai, M.R. Morris, and  G. Villar, Astron. J. {\bf 141},  134  (2011)

\item O. Struve and P. Swings, Astrophys. J. {\bf 93}, 356 (1941) 

\item O.G. Taranova, V.I. Shenavrin, Astron. Lett. {\bf 33}, 584 (2007) 

\item B.A. Vorontsov-Velyaminov, Sov. Astron. {\bf 4}, 929 (1960) 

\item B.A. Vorontsov-Velyaminov, E.B. Kostyakova, O.D. Dokuchaeva, V.P. Arkhipova, Sov. Astron. {\bf 9}, 364 (1965)

\item B.A. Vorontsov-Velyaminov, E.B. Kostyakova, O.D. Dokuchaeva, V.P. Arkhipova, Trudy GAISh {\bf 40}, 57 (1970)

\item M.L. White, Astrophys. J. {\bf 115}, 71 (1952)

\item P.A. Whitelock, MNRAS {\bf 213}, 59 (1985)

\end{enumerate}


\begin{center}
\begin{longtable}{@{}cccccc}
\caption{$UBV$ photometry for IC~4997 in 2009-2019.}\\
\label{ubv} \\
\hline
JD&$V$&$B$&$U$\\
\hline
\endfirsthead
\multicolumn{4}{c}%
 {\tablename\ \thetable\ -- \textit{}} \\ \hline
  JD&$V$&$B$&$U$\\
 \hline \\
\endhead
\multicolumn{6}{r}{\textit{}} \\ \hline
\endfoot
\hline
\endlastfoot

2455031 &10.552&11.200&10.770\\
2455038 &10.549&11.213&10.768\\
2455042 &10.578&11.204&10.786\\
2455057 &10.575&11.197&10.810\\
2455061 &10.579&11.210&10.761\\
2455068 &10.561&11.214&10.771\\
2455086 &10.572&11.183&10.716\\
2455092 &10.582&11.186&10.763\\
2455095 &10.604&11.212&10.775\\
2455331 &10.601&11.214&10.719\\
2455362 &10.543&11.204&10.784\\
2455363 &10.529&11.192&10.772\\
2455399 &10.528&11.207&10.830\\
2455410 &10.495&11.200&10.744\\
2455413 &10.489&11.202&10.769\\
2455422 &10.519&11.205&10.787\\
2455662 &10.543&11.204&10.784\\
2455663 &10.529&11.192&10.772\\
2455737 &10.544&11.194&10.772\\
2455743 &10.565&11.200&10.784\\
2455750 &10.569&11.192&10.766\\
2455754 &10.531&11.190&10.792\\
2455766 &10.535&11.209&10.765\\
2455774 &10.529&11.206&10.767\\
2455779 &10.554&11.204&10.767\\
2455780 &10.650&11.248&10.811\\
2455782 &10.542&11.188&10.771\\
2455794 &10.552&11.205&10.777\\
2455861 &10.570&11.211&10.750\\
2455866 &10.598&11.215&10.782\\
2456101 &10.515&11.196&10.767\\
2456121 &10.484&11.184&10.791\\
2456122 &10.531&11.186&10.780\\
2456123 &10.511&11.187&10.779\\
2456130 &10.483&11.191&10.776\\
2456153 &10.561&11.198&10.783\\
2456155 &10.563&11.198&10.774\\
2456161 &10.534&11.200&10.757\\
2456448 &10.520&11.188&10.760\\
2456457 &10.514&11.188&10.789\\
2456464 &10.505&11.189&10.779\\
2456483 &10.522&11.190&10.772\\
2456504 &10.518&11.188&10.764\\
2456514 &10.495&11.183&10.775\\
2456518 &10.503&11.188&10.797\\
2456575 &10.552&11.202&10.778\\
2456605 &10.553&11.199&10.780\\
2456607 &10.540&11.200&10.765\\
2456839 &10.486&11.131&10.765\\
2456868 &10.477&11.132&10.765\\
2456875 &10.468&11.108&10.744\\
2456885 &10.442&11.147&10.715\\
2456893 &10.506&11.123&10.775\\
2456944 &10.547&11.170&10.746\\
2457213 &10.484&11.130&10.827\\
2457216 &10.478&11.138&10.778\\
2457220 &10.497&11.136&10.778\\
2457253 &10.501&11.183&10.810\\
2457270 &10.447&11.169&10.767\\
2457550 &10.487&11.200&10.774\\
2457578 &10.506&11.128&10.768\\
2457583 &10.457&11.142&10.784\\
2457640 &10.442&11.210&10.694\\
2457935 &10.442&11.196&10.742\\
2457950 &10.478&11.195&10.749\\
2457958 &10.487&11.189&10.740\\
2457959 &10.481&11.166&10.781\\
2457967 &10.440&11.199&10.751\\
2457969 &10.386&11.190&10.750\\
2457979 &10.416&11.124&10.766\\
2457986 &10.448&11.131&10.775\\
2457994 &10.461&11.126&10.766\\
2458013 &10.454&11.182&10.763\\
2458046 &10.564&11.192&10.761\\
2458282 &10.449&11.187&10.754\\
2458306 &10.499&11.193&10.786\\
2458347 &10.418&11.182&10.770\\
2458364 &10.471&11.121&10.773\\
2458435 &10.474&11.220&10.760\\
2458613	&10.490&11.191&10.773\\
2458636	&10.425&11.141&10.789\\
2458638	&10.434&11.119&10.785\\
2458647	&10.445&11.123&10.756\\
2458658	&10.437&11.120&10.781\\
2458661	&10.449&11.130&10.776\\
2458677 &10.443&11.132&10.758\\
2458691	&10.440&11.111&10.743\\
2458701	&10.468&11.118&10.778\\
2458704	&10.441&11.124&10.768\\
2458719	&10.388&11.103&10.794\\
2458720	&10.415&11.119&10.727\\
2458721	&10.399&11.112&10.743\\
2458725	&10.446&11.126&10.804\\
2458733	&10.472&11.101&10.761\\
2458752 &10.477&11.119&10.774\\
2458760	&10.485&11.107&10.797\\
2458778	&10.472&11.127&10.795\\
2458789	&10.492&11.129&10.787\\

\end{longtable}
\end{center}

\begin{table}
 \caption{The annual average $UBV$ magnitudes for IC~4997 in 2009-2019.}\label{mean}
 \begin{center}
 \begin{tabular}{cccccccc}
  \hline
Year & $V$&$\sigma_V$& $B$ &$\sigma_B$& $U$&$\sigma_U$&$N$\\

  \hline
2008.6&10.572&0.017&11.202&0.012&10.769&0.012&10\\
2009.5&10.529&0.037&11.204&0.007&10.772&0.035&7\\
2011.6&10.523&0.031&11.192&0.006&10.776&0.010&8\\
2010.6&10.558&0.035&11.204&0.015&10.776&0.015&14\\
2012.6&10.523&0.020&11.191&0.006&10.776&0.011&10\\
2013.6&10.488&0.033&11.135&0.021&10.754&0.020&7\\
2014.6&10.481&0.021&11.151&0.023&10.792&0.025&5\\
2015.6&10.473&0.029&11.170&0.041&10.755&0.041&4\\
2016.6&10.460&0.045&11.172&0.030&10.759&0.013&11\\
2017.6&10.462&0.030&11.180&0.037&10.769&0.012&5\\
2018.6&10.447&0.033&11.127&0.021&10.774&0.021&19\\

 \hline
 \end{tabular}
 \end{center}
\end{table}


\begin{table}
 \caption{$JHKL$ photometry for IC~4997 in 2019.}\label{IR}
 \begin{center}
 \begin{tabular}{ccccccccc}
  \hline
JD& $J$&$\sigma_J$& $H$ &$\sigma_H$& $K$&$\sigma_K$&$L$&$\sigma_L$\\

  \hline
2458631.5&	10.52&	0.03&	10.67&	0.05&	9.94&	0.03&	--&	--\\
2458634.5&	10.42&	0.04&	10.60&	0.04&	9.91&	0.03&	8.72&	0.09\\
2458631.5&	10.52&	0.03&	10.67&	0.05&	9.94&	0.03&	--&	--\\
2458655.5&  10.51&	0.02&	--   &	--	&   9.92&	0.03&   8.70&--\\
2458659.5&	10.44&	0.04&	10.65&	0.07&	9.92&	0.03&	8.70&0.08\\
2458682.5&	10.53&	0.03&	10.69&	0.05&	9.99&	0.04&	8.54&	0.12\\
2458686.5&	10.52&	0.03&	10.71&	0.04&	9.93&	0.03&	8.52&	0.11\\
2458704.5&	10.55&	0.04&	10.52&	0.06&	9.93&	0.05&	8.48&	0.18\\
2458734.3&  10.54&  0.03&   10.52&  0.03&   9.94&   0.03&   8.56&   0.08\\
2458776.2&  10.47&  0.03&   10.64&  0.03&   9.95&   0.02&   8.61&   0.10\\ 

 \hline
\end{tabular}
\end{center}
\end{table}


\begin{table}
\caption{Log of spectroscopic observations of IC~4997.}\label{sp_log}
    \begin{center}
    \begin{tabular}{cccc}
    \hline
         Date & JD & Range, \AA & Standard \\
         \hline
         04.08.2010 & 2455413 & 4000--7200 & 18~Vul \\
         31.07.2011 & 2455774 & 4000--7400 & HD~196775 \\
         26.08.2011 & 2455800 & 4000--9100 & 40~Cyg \\
         21.07.2012 & 2456130 & 4000-7400 & 18~Vul \\
         23.08.2012 & 2456163 & 4200--7400 & $\rho$~Aql \\
         09.07.2013 & 2456483 & 4000--7200 & 29~Vul \\
         08.08.2015 & 2457243 & 4000--7700 & 107~Her \\
         06.08.2016 & 2457607 & 4000--6700 & $\rho$~Aql \\
         09.08.2016 & 2457610 & 4000--7700 & $\rho$~Aql \\
         21.07.2017 & 2457956 & 4000--9700 & $\rho$~Aql \\
         19.10.2017 & 2458046 & 4000--9700 & $\rho$~Aql \\
         08.08.2018 & 2458339 & 4000--7000 & 29~Vul \\
         13.10.2018 & 2458405 & 4000--9400 & 29~Vul \\
         07.07.2019 & 2458672 & 4000--9400 & 29~Vul, 30~Vul \\ 
         07.08.2019 & 2458703 & 4000--9250 & 29~Vul, HD~196775, $\eta$~Sge \\
         03.10.2019 & 2458760 & 4000--9250 & HD~196775, $\eta$~Sge \\
    \hline
    \end{tabular}
    \end{center}
\end{table}


\begin{table}
    \caption{The observed relative intensities of emission lines for IC~4997 on the scale $I(\textrm{H}\beta)=100$ and the observed intensity of $\textrm{H}\beta$ in units of 10$^{-11}$ erg cm$^{-2}$s$^{-1}$.}\label{intensity}
    \begin{center}
    \begin{tabular}{cccccccccc}
    \hline   
    $\lambda$, \AA & Ion & 04.08.10 & 31.07.11 & 26.08.11 & 21.07.12 & 23.08.12 & 09.07.13 & 08.08.15 & 06.08.16\\
    \hline
    4102 & H$\delta$ & 20.1 & 20.7 & 20.9 & 23.8 & - & 21.6 & 21.9 & 21.4\\
    4340 & H$\gamma$ & 43.7 & 43.7 & 46.6 & 51.0 & 46.5 & 46.4 & 47.9 & 42.2\\
    4363 & [OIII] & 46.3 & 46.8 & 41.8 & 43.5 & 43.9 & 41.6 & 36.8 & 33.9\\
    4471 & HeI & 4.01 & 4.35 & 4.47 & 4.81 & 4.13 & 4.54 & 4.45 & 4.16\\
    4713 & HeI & 0.94 & 0.97 & 0.85 & 0.97 & - & 0.97 & 0.81 & 0.76\\
    4740 & [ArIV] & 0.38 & 0.45 & 0.41 & 0.43 & - & 0.42 & 0.37 & 0.34\\
    4959 & [OIII] & 227 & 238 & 238 & 224 & 233 & 237 & 225 & 226\\
    5007 & [OIII] & 690 & 725 & 734 & 680 & 691 & 723 & 678 & - \\
    5192 & [ArIII] & 0.18 & 0.16 & 0.20 & 0.16 & - & - & 0.11 & 0.11\\
    5270 & [FeIII] & 0.27 & 0.30 & 0.27 & 0.32 & - & 0.28 & 0.26 & 0.29\\
    5518 & [ClIII]+OI & 0.18 & 0.15 & 0.14 & 0.15 & - & 0.17 & 0.14 & 0.14\\
    5538 & [ClIII] & 0.28 & 0.29 & 0.28 & 0.31 & - & 0.33 & 0.30 & 0.29\\
    5755 & [NII] & 1.45 & 1.65 & 1.66 & 1.56 & 1.64 & 1.60 & 1.50 & 1.48\\
    5876 & HeI & 20.8 & 21.3 & 21.2 & 20.4 & 21.5 & 16.9 & 19.1 & 19.4\\
    6300 & [OI] & 6.82 & 5.91 & 6.73 & 6.35 & 6.70 & 6.38 & 5.88 & 6.10\\
    6312 & [SIII] & 3.03 & 2.93 & 3.25 & 3.40 & 3.13 & 2.95 & 3.12 & 3.09\\
    6364 & [OI] & 2.29 & 2.16 & 2.33 & 2.27 & 2.24 & 2.23 & 2.19 & 2.25\\
    6563 & H$\alpha$ & 376 & 397 & 380 & 366 & 463 & 376 & 425 & - \\
    6584 & [NII] & 24.7 & 26.2 & 28.1 & 25.4 & 33.3 & 27.4 & 30.1 & 38.8\\
    6678 & HeI & 5.52 & 5.75 & 5.69 & 5.41 & 6.56 & 5.11 & 5.31 & - \\
    6716 & [SII] & 0.86 & 0.96 & 0.92 & 0.88 & 1.14 & 0.77 & 1.04 & - \\
    6731 & [SII] & 1.85 & 1.95 & 1.91 & 1.85 & 2.34 & 1.71 & 1.89 & - \\
    7065 & HeI & 17.0 & 19.5 & 18.5 & 18.6 & 20.9 & 15.8 & 17.6 & - \\
    7136 & [ArIII] & 12.6 & 15.1 & 13.7 & 14.9 & 16.7 & 12.4 & 14.4 & - \\
    7170 & [ArIV] & 0.61 & 0.35 & 0.54 & 0.34 & - & - & 0.62 & - \\
    7237 & [ArIV] & - & 0.26 & 0.32 & 0.28 & - & - & 0.44 & - \\
    7281 & HeI & - & 1.63 & 1.39 & 1.31 & 1.18 & - & 1.70 & - \\
    7751 & [ArIII] & - & - & 3.12 & - & - & - & - & - \\
    9069 & [SIII] & - & - & 21.7 & - & - & - & - & - \\
    \hline
     & $F(\textrm{H}\beta)$  & 3.14 & 2.87 & 3.24 & 3.05 & 3.04 & 3.24 & 3.25 & 3.33\\
    \hline
    \end{tabular}
    \end{center}
\end{table}
\addtocounter{table}{-1}
\begin{table}
    \caption{Continued}
    \begin{center}
    \begin{tabular}{cccccccccc}
    \hline
    $\lambda$, \AA & Ion & 09.08.16 & 21.07.17 & 19.10.17 & 08.08.18 & 13.10.18 & 07.07.19 & 07.08.19 & 03.10.19 \\
\hline 
4102 & H$\delta$ & 18.2 & 19.6 & 19.7 & 19.9 & 17.4 & 19.3 & 19.8 & 19.4\\
4340 & H$\gamma$ & 41.2 & 45.7 & 43.1 & 46.8 & 42.3 & 42.8 & 42 & 40\\
4363 & [OIII] & 30.6 & 32.2 & 30.8 & 32.5 & 29.0 & 32.6 & 35.5 & 32.7\\
4471 & HeI & 3.78 & 4.01 & 3.53 & 4.06 & 3.16& 3.39 & 4.79 & 4.51\\
4712 & HeI  & 0.68 & 0.83 & 0.73 & 0.77 & 0.77 & 0.92 & 1.05 & 0.9\\
4740 & [ArIV] & 0.32 & 0.36 & 0.32 & 0.29 & 0.30 & 0.44 & 0.49 & 0.46\\
4959 & [OIII] & 226 & 219 & 225 & 232 & 235 & 230 & 237 & 216\\
5007 & [OIII] & - & 671 & 670 & - & 715 & - & 726 & 658 \\
5192 & [ArIII] & 0.11 & 0.13 & - & 0.13 & 0.10 & 0.10 & 0.15 & 0.09\\
5270 & [FeIII] & 0.27 & 0.26 & - & 0.27 & 0.26 & 0.26 & 0.27 & 0.31\\
5518 & [ClIII]+OI & 0.13 & 0.12 & 0.14 & 0.12 & 0.16 & 0.18 & 0.15 & 0.16\\
5538 & [ClIII] & 0.28 & 0.29 & 0.25 & 0.30 & 0.31 & 0.33 & 0.35 & 0.33\\
5755 & [NII] & 1.53 & 1.39 & 1.37 & 1.41 & 1.34 & 1.34 & 1.66 & 1.61\\
5876 & HeI & 19.1 & 18.9 & 16.6 & 18.7 & 18.9 & 18.1 & 19.4 & 19.7\\
6300 & [OI] & 6.25 & 6.43 & 5.33 & 5.77 & 6.01 & 5.8 & 7.16 & 6.94\\
6312 & [SIII] & 3.00 & 2.58 & 2.63 & 2.87 & 2.81 & 2.48 & 2.79 & 2.87\\
6364 & [OI] & 2.29 & 2.23 & 1.62 & 2.05 & 2.16 & 2.11 & 2.44 & 2.48\\
6563 & H$\alpha$ & 351: & 438 & 321: & 422 & 388 & 368 & 475 & 408\\
6584 & [NII] & 35.7 & 38.6 & 27.5 & 36.7 & 36.2 & 33.7 & 32.2 & 28.4\\
6678 & HeI & 5.03 & 5.33 & 4.12 & 4.88 & 5.01 & 4.92 & 5.48 & 5.75\\
6716 & [SII] & 1.02 & 0.85 & 0.73 & 0.75 & 0.84 & 0.88 & 0.97 & 0.97\\
6731 & [SII] & 1.77 & 1.93 & 1.50 & 1.79 & 1.84 & 1.82 & 2.08 & 2.20\\
7065 & HeI & 16.2 & 16.7 & 13.9 & - & 15.1 & 13.9 & 16.0 & 18.5\\
7136 & [ArIII] & 13.5 & 13.8 & 11.1 & - & 12.2 & 11.4 & 13.3 & 14.8\\
7170 & [ArIV] & - & 0.47 & 0.43 & - & 0.40 & 0.47  & 0.40 & 0.63\\
7237 & [ArIV] & - & 0.35 & 0.28 & - & 0.33 & 0.29 & 0.40 & 0.35\\
7281 & HeI & 1.10 & 1.80 & 1.29 & - & 1.68 & 1.43 & 1.57 & 1.74\\ 
7751 & [ArIII] & - & 3.63 & 2.17 & - & 3.12 & 2.85 & 3.46 & 3.86\\
9069 & [SIII] & - & 43.9 & 36.6 & - & 26.2 & 32.3& 57.9 & 46.2\\
\hline
     $F(\textrm{H}\beta)$ & 3.33 & 3.55 & 3.64 & 4.29 & 3.08 & 3.19 & 3.53 & 2.93 & 3.16\\
\hline
    \end{tabular}
    \end{center}
\end{table}


\begin{table}
\caption{Simulated physical conditions in the zone radiating H$\beta$, [OIII] $\lambda4363,4959$ for the 1970-2019 period.}\label{NeTe}
    \begin{center}
    \begin{tabular}{lcccc}
    \hline
         & 1970 & 1990 & 2003 & 2019\\
         \hline
         $T_e$,~K& 8000 & 11900 & 9740 & 7650 \\
         $N_e$,~см$^{-3}$ & 5.6(6) & 5.3(6) & 5.0(6) & 5.14(6)\\
         $\rm O^{2+}/\rm H^+$ & 4.2(-3) & 6.5(-4) & 2.18(-3) & 6.0(-3)\\
         $F(\lambda5876)/F(\textrm{H}\beta)$ & 0.19 & 0.22 & 0.20 & 0.19\\
         \hline
         $T_e$,~K& 10000 & 16700 & 12800 & 9440 \\
         $N_e$,~см$^{-3}$ & 2.16(6) & 2.3(6) & 2.05(6) & 2.01(6)\\
         $\rm O^{2+}/\rm H^+$ & 7.7(-4) & 1.27(-4) & 2.4(-4) & 1.13(-3)\\
         $F(\lambda5876)/F(\textrm{H}\beta)$ & 0.20 & 0.29 & 0.23 & 0.20\\
         \hline
         $T_e$,~K& 12000 & 22600 & 16150 & 12220 \\
         $N_e$,~см$^{-3}$ & 1.18(6) & 1.4(6) & 1.17(6) & 1.08(6)\\
         $\rm O^{2+}/\rm H^+$ & 2.7(-4) & 4.6(-5) & 1.5(-4) & 4.0(-4)\\
         $F(\lambda5876)/F(\textrm{H}\beta)$ & 0.22 & 0.36 & 0.28 & 0.21\\
         \hline
    \end{tabular}
    \end{center}
\end{table}

\end{document}